\documentclass[runningheads]{llncs}
\usepackage{graphicx}
\usepackage{multirow}
\usepackage{amsmath}
\usepackage{xcolor}

\let\llncssubparagraph\subparagraph
\let\subparagraph\paragraph
\usepackage[compact]{titlesec}
\let\subparagraph\llncssubparagraph

\usepackage{caption}
\newcommand\blfootnote[1]{%
  \begingroup
  \renewcommand\thefootnote{}\footnote{#1}%
  \addtocounter{footnote}{-1}%
  \endgroup
}
\begin{document}
\title{Detecting Changes in Crowdsourced Social Media Images}
%
%

\author{}
\institute{}
\author{Muhammad Umair
\and
Athman Bouguettaya
\and
Abdallah Lakhdari
} 
%
\authorrunning{Umair et al.}
%
\institute{The University of Sydney, Sydney, NSW 2006, Australia \\
\email\{muhammad.umair, athman.bouguettaya, abdallah.lakhdari\}@sydney.edu.au}%
\maketitle              
\begin{abstract}
We propose a novel service framework to detect changes in crowdsourced images. We use a service-oriented approach to model and represent crowdsourced images as \textit{image services}. \textit{Non-functional attributes} of an image service are leveraged to detect changes in an image. 
The changes are reported in form of a \textit{version tree}. The version tree is constructed in a way that it reflects the extent of changes introduced in different versions. Afterwards, we find semantic differences in between different versions to determine the extent of changes introduced in a specific version. Preliminary experimental results demonstrate the effectiveness of the proposed approach.
\keywords{Image as a service \and Modified images \and Version tree \and Fake images \and Fake news \and Trust \and Image provenance \and Social media \and Big data
}
\end{abstract}
\blfootnote{Acknowledgment: This research was partly made possible by LE220100078 and DP220101823 grants from the Australian Research Council. The statements made herein are solely the responsibility of the authors.}


\section{Introduction}\label{sec:into}

Social media has become a key platform to share news and information related to public incidents~\cite{ali2017sentiment}. There are more than 5 billion active users on social media~\cite{griffis2014use}. Social media users publish a large amount of data related to public events~\cite{liu2011using}. These social media images may contain critical information about public incidents i.e., road accidents, crime scenes, violent scenes, etc. The images related to a particular incident may have different versions uploaded on social media. Utilizing these versions can significantly facilitate the task of scene reconstruction to explore unfolding situations which might have led to the incident.

Existing work on scene reconstruction is based on image processing that is usually computationally intensive~\cite{chae2012spatiotemporal}. 
A novel technique has recently been proposed to reconstruct scenes using images' metadata~\cite{aamir2020heuristics}. It leverages the service paradigm to represent social media images and related posted information as \textit{services}.  It abstracts social media users as \textit{social-sensors} and an image as a \textit{social-sensor service}, abbreviated as \textit{SocSen}.
Henceforth, we use the term `\textit{image service}' to refer to `\textit{social-sensor service}'. The only difference between an image service and a SocSen service is that an image service comparatively contains a vast set of non-functional attributes.
An image service is defined to have \textit{functional} and \textit{non-functional} properties. Functional attributes are the parameters related to the actions pertaining to the capture of an image service. Examples of functional attributes are switching picture/video modes, pressing on/off button, delayed/timed picture taking, taking panoramic shots, etc. Non-functional attributes facilitate the delivery of the purpose of taking a picture. Examples of non-functional attributes are \textit{subject distance}, \textit{camera elevation angle}, \textit{resolution}, \textit{location} etc. The non-functional attributes are usually available in form of different \textit{metadata} tags. An image service can also have different versions. 

Most existing work on image services is related to image service selection and composition~\cite{aamir2020social,mistry2016metaheuristic}. 
An image service composition approach has been proposed to form a tapestry in the spatial aspect and a storyboard in the temporal aspect~\cite{aamir2020heuristics}. 
The focus of most existing work on image services has hitherto been to reconstruct a scene. A fundamental assumption in this regard has so far been that the participating image services are intrinsically \textit{trustworthy}. However, the \textit{trust} issue becomes paramount when image services are assumed to be crowdsourced~\cite{shen2019fake}. For instance, a crime scene analysis relying on crowdsourced image services may contain untrustworthy images which may lead to wrong conclusions. Untrustworthy image services can be avoided by analyzing different versions of a social media image to find the most trustworthy version for the scene reconstruction.

Detecting untrustworthy image services has traditionally been addressed using image processing and information retrieval techniques~\cite{gupta2013faking}. These approaches are usually costly and computationally expensive. A preliminary service-based trust framework is proposed in~\cite{aamir2018trust,aamir2018stance} which is based on users' comments and stances in an image service to assess its credibility. 
However, the credibility of image services may not be completely assessed based only on the user's stance. \textit{Fake} posts on social media can get supportive comments from other users~\cite{colliander2019fake}. The stance of credible users may also be biased~\cite{zimmer2019fake}. Moreover, these approaches focus on changes in an individual image service. Whereas, multiple social media images are being forged in conjunction to manipulate information about an incident. To address these limitations, \textit{we propose to assess trust among image services using a more objective and holistic framework consisting of changes and updates in different versions of the image services.} 
In this regard, we leverage non-functional attributes of different versions of an image to detect changes in image services. 
\textit{we operate under the assumption that the non-functional attributes, encompassing critical technical details, can be seamlessly accessed through the image service}.

We propose a novel approach to detect changes in the non-functional attributes of an image service, as a first step towards ascertaining whether an image service is fake. Editing an image service may make it inconsistent with its non-functional attributes. \color{black}An attempt to hide the facts in metadata may create some discrepancies among non-functional attributes. These discrepancies are not straightforward to identify as they are usually embedded in the non-functional attributes. 
We utilize these inconsistencies to detect changes in an image service.
In this respect, we form different groups of image non-functional attributes such that analyzing each group collectively provides useful insights on inconsistencies. For instance, \textit{shutter speed}, \textit{exposure time} and \textit{aperture size} are the attributes that inform about the intake of light while capturing an image. Analyzing them collectively may indirectly reflect the \textit{time of day} while the image service was captured. Changes in different versions of an image service are also investigated to get more insights about the transformation of different versions from the original image. In this regard, we assume that the provided image service is not the only version of an image upload on social media. We use an image-based search i.e., Reverse Image Search (RIS) to collect all versions of an image service available on social media. A temporal sorting is then performed to arrange them in a sequence they were uploaded on social media. We propose a novel representation of changes in an image service in terms of a \textit{version tree}. The version tree is constructed in a way such that the information about changes in a specific version is implicit in its position/placement in the tree. 
Therefore, one of the main contribution of this paper is to propose a framework to build the version tree.
Afterwards, a state-of-the-art semantic similarity measure is used to find semantic differences between an image and its versions. The proposed framework is a kind of image provenance analysis based only on the non-functional attributes. The proposed approach is validated on 5849 images collected from an image metadata dataset.  
Below, we summarize our main contributions:
\begin{itemize}
    \item A framework is proposed to detect changes in different versions of an image service using only the non-functional attributes.
    \item We introduce a unique way of reporting changes in different versions of an image service in terms of a version tree. Knowledge about changes in all versions is implicit in the tree.
    \item The proposed framework also provides a serendipitous image provenance analysis using the version tree. 
\end{itemize}

The proposed method effectively handles a typical set of modifications that are reflected in the non-functional attributes. Although, the non-functional attributes may not completely capture certain changes within the image itself, such as alterations in shades, intensity of colors, or distortion, it remains well-suited for a wide range of image modifications. It is worth noting that the proposed framework's performance may be influenced by the availability of non-functional attributes. In cases where a limited number of such attributes are available, an alternative approach involves obtaining meta-information from the social media post. While this alternative approach may be less precise due to potential questions about the accuracy of meta-information, it still offers valuable insights. 


\section{Motivating Scenario}\label{sec:motivation}

\begin{figure}
\vspace{-.5em}
\centerline{\includegraphics[width = 1.05\columnwidth]{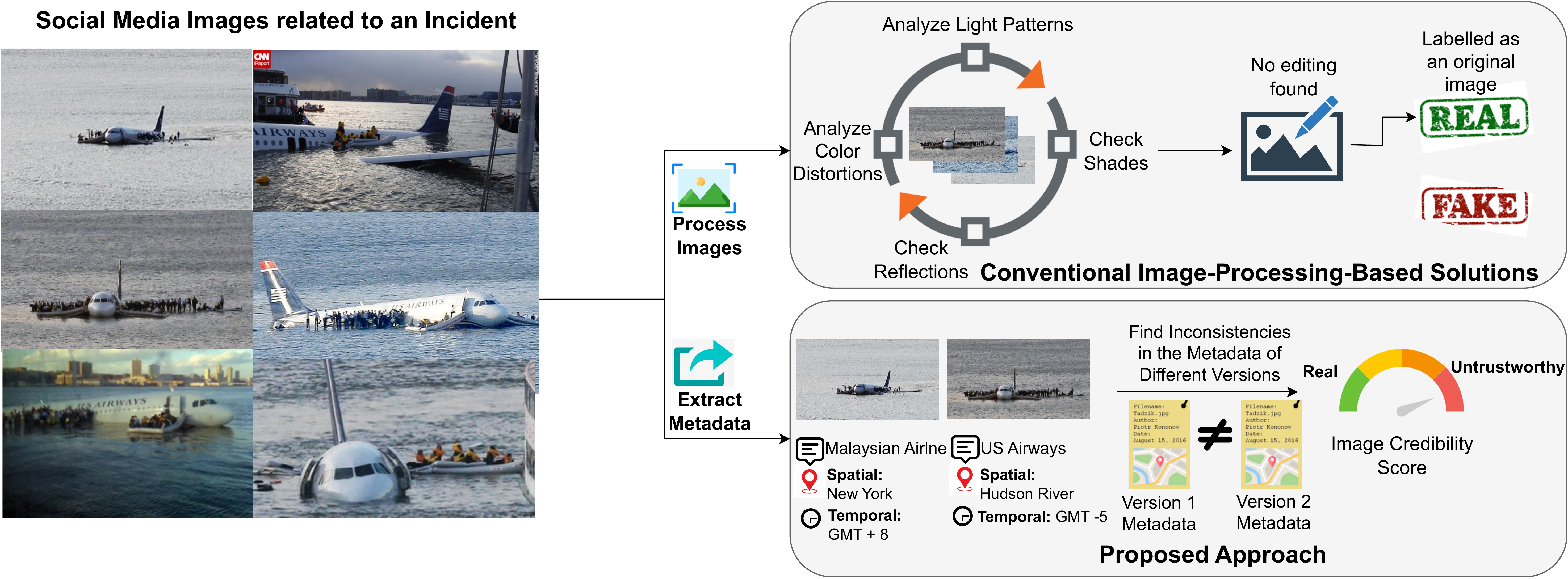}}
\caption{Motivating Scenario} \label{fig:motivation}
\vspace{-3.5em}
\end{figure}

We consider a scene of a plane crash in New York that happened in 2009 as our motivating scenario. Figure~\ref{fig:motivation} shows evacuation of US Airways Flight 1549 as it floats on the Hudson River. This image was falsely claimed as the lost Malaysian aircraft MH370 in many social media posts in 2014. In those misleading posts, the images are original but they contain a false claim. Many state-of-the-art solutions rely on image processing to identify untrustworthy social media images~\cite{shahzad2022review}. These solutions focus on the content within the images and hence, they may fail to identify changes which are not in images. Moreover, these solutions consider changes in a single image and provide no knowledge about the image provenance. To address these limitations, we propose a unique way of identifying untrustworthy images by doing image provenance analysis using only the non-functional attributes of an image and its different versions. Exploring the non-functional attributes of different versions of an image may reveal a lot of inconsistencies among them. These inconsistencies reflect the \textit{trust} of an image. For instance, in Figure~\ref{fig:motivation}, metadata of different versions of an image is inconsistent and is reflective of the modifications in different versions. 

\section{Related Work}\label{sec:related}
Existing work on image services is related to image service selection and composition strategies. 
A context and direction aware spatio-temporal clustering is proposed in \cite{aamir2020social}. The proposed approach helps to compose the relevant images to form a tapestry in the spatial aspect and a story in the temporal aspect.
A fundamental assumption in this work has hitherto been that images participating in scene reconstruction are trustworthy. However, image services in a crowdsourced environment can be untrustworthy. Traditional approaches to identify untrustworthy images are based on image processing and machine learning~\cite{patel2022fake}.
Moreover, some text classification techniques are also available in the literature that can be employed to classify fake text with an image~\cite{umair2020multi}.
The aforementioned image processing based technique has high accuracies in determining the fakes in an image but require high computational power.

Some recent studies claim that a subset of \textit{trust} can be derived using light weight service-oriented approaches~\cite{aamir2018stance,aamir2018trust}. A new image services \textit{trust} model is proposed in~\cite{aamir2018trust}. The trustworthiness of an image service is measured based on the users' stance. Textual features of the image services, i.e., comments are utilized to determine the trust of the service. Another users’ stance and credibility-based image service's trust model is proposed in~\cite{aamir2018stance}. The proposed model considers various indicators such as the stance embedded in the services’ comments, their meta-data, e.g., time, along with the users’ credibility. 
These approaches are unable to capture modifications in an image service because the misleading content on social media may receive positive comments from other users~\cite{colliander2019fake}. Moreover, comments from credible users can be biased.

We propose a relatively more objective and holistic approach that considers modifications and updates introduced in an image service to determine the trust of an image service. This paper focuses on detecting the changes, as a first step towards determining the trust. In this regard, different versions of an image are investigated to get more insights on image provenance details. Versions of an image are being utilized in many state-of-the-art solutions to identify changes in an image. For instance, an image provenance analysis is proposed in \cite{bharati2019beyond} using the metadata. Different versions of an image are investigated in \cite{saez2014fake} to explore different contexts in which the image was shared. Different versions of video clips are analyzed in \cite{ganti2022novel} to detect misinformation in videos.
A framework is proposed in \cite{qian2023fighting} that relies on different versions of images to determine if they are shared out of context. The difference between these approaches and our proposed framework is that these approaches majorly rely on computer vision to determine fake images, whereas, our proposed approach is completely based on analyzing metadata of different versions to determine changes in an image.
It is worth clarifying that the proposed framework is different from object versioning.  
The focus of object versioning systems is the creation and management of versions. Whereas, the focus of the proposed framework is to find image versions based on the changes in the non-functional attributes of an image service.


\section{Image Service Model}\label{sec:ImageServiceModel}

We represent an image service in terms of its functional and non-functional attributes:
\begin{equation}
    ImgServ = \{f\} \cup \{nf\} 
    \label{eq1}
\end{equation}
where \textit{f} and \textit{nf} represent the set of functional and non-functional attributes respectively. Functional attributes represent the actions involved in capturing an image. Functional attributes can be formalized as:
 \begin{equation}
     f = \{\alpha, \mu, \gamma\} 
     \label{eq2}
 \end{equation}
where $\alpha$ represents the action of capturing an image, i.e., pressing the shutter, $\mu$ is the action to switch from one capturing mode to another i.e., switching from picture to video and vice versa, and $\gamma$ represents the time delay in taking a picture.
Non-functional part of an image service consists of spatio-temporal, contextual and intrinsic attributes as listed in Table~\ref{FeatureVariables}.
\begin{equation}
     nf = \{\zeta, \tau, c, \iota\} 
     \label{eq2}
\end{equation}

where $\zeta$ represents the set of spatial attributes, $\tau$ is the set of temporal attributes, \textit{c} contains contextual attributes, and $\iota$ represents intrinsic attributes. Intrinsic attributes reflect changes inside the images i.e., changes in the visual content. 


\begin{table}
\caption{Description of Non-functional Attributes}\label{FeatureVariables}
\begin{tabular}{|p{2cm}|p{6cm}|l|}
\hline
\textbf{Categories} & \textbf{
Description} & \textbf{Example Attributes}\\
\hline
\multirow{2}{2cm}{Spatial Features}\label{sec:Non-functionalAttributes}
& \multirow{2}{6cm}{Spatial metadata tags describe the location at which the image was taken.} & GPS Coordinates \\  
  &&City, Sate, Country \\  
\hline  
\multirow{2}{2cm}{Temporal Features}
& \multirow{2}{6cm}{Temporal metadata tags describe the date and time when the image was taken.}  
  & GPS Timezone Offset \\ &&GPS Timestamp  \\  
\hline  
\multirow{3}{2cm}{Contextual Features}
& \multirow{2}{6cm}{Contextual features define the context of an image. Contextual attributes may also contain 
the details of the ambiance.} & Title \\  
  &&Caption \\ &&Headline \\ 
\hline
\end{tabular}
\end{table}

\noindent\textbf{\underline{Non-functional Attributes of an Image Service:}}
We identify different non-functional attributes of an image service that may indicate changes within an image (refer to Table~\ref{FeatureVariables}).
We group the non-functional attributes into the following categories:

\begin{itemize}

    \item \textit{Spatial Features}: Spatial attributes represent the location where the image was captured. Modified spatial tags may be an indication of fake background.
    
    \item \textit{Temporal Features}: Temporal attributes represent the date and time when the image was captured. 
    Forged temporal metadata tags develop a fake story.
    
    \item \textit{Contextual Features}: Contextual features are related to the context of an image. Fake context may support fake spatio-temporal tags of an image.
\end{itemize}


\noindent\textbf{\underline{Potential Modifications in Non-functional Attributes:}}\label{sec:Modifications}
We propose a categorization of potential modifications that may exist in an image service. 
The following are the possible changes in image's non-functional attributes:

\begin{figure}
\vspace{-1.5em}
\centerline{\includegraphics[width = .55\columnwidth]{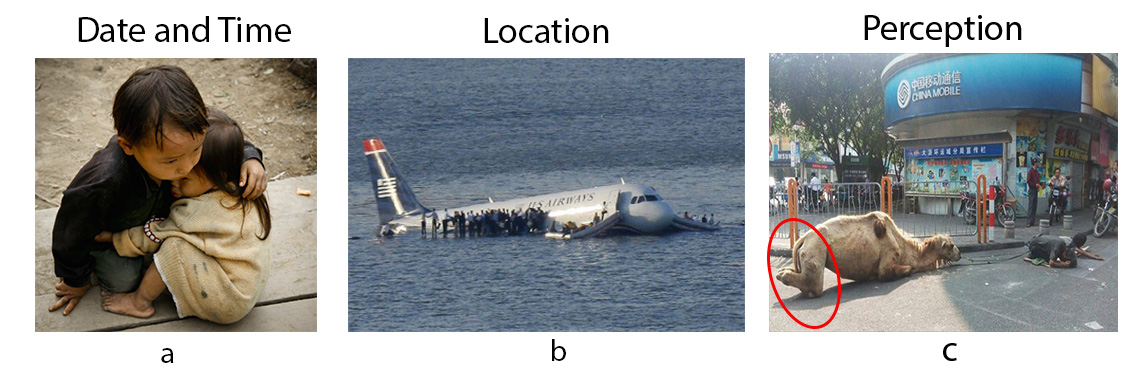}}
\caption{Potential Changes in Images' Non-functional Attributes}
\vspace{-2.5em}
\label{fig:ext_changes}
\end{figure}

\begin{itemize}
    \item \textit{Modified Date and Time}:
    Figure~\ref{fig:ext_changes}a claims two children in 2015 Nepal earthquake. It is actually a picture of two Vietnamese taken in 2007.
    
    \item \textit{Modified Location}:
    Figure~\ref{fig:ext_changes}b was a viral photo in 2014 claiming the picture of the lost Malaysian MH370 plane. It turned out to be a photo of a plane crash in New York in 2009.
    
    \item \textit{Modified Context}: 
    Figure~\ref{fig:ext_changes}c claims a camel with limbs cut off used for begging. The camel is actually resting with legs bent under itself.
    
\end{itemize}


\section{Proposed Framework}\label{sec:framework}
\begin{figure}
\vspace{-1.5em}
\centerline{\includegraphics[width = 1.1\columnwidth]{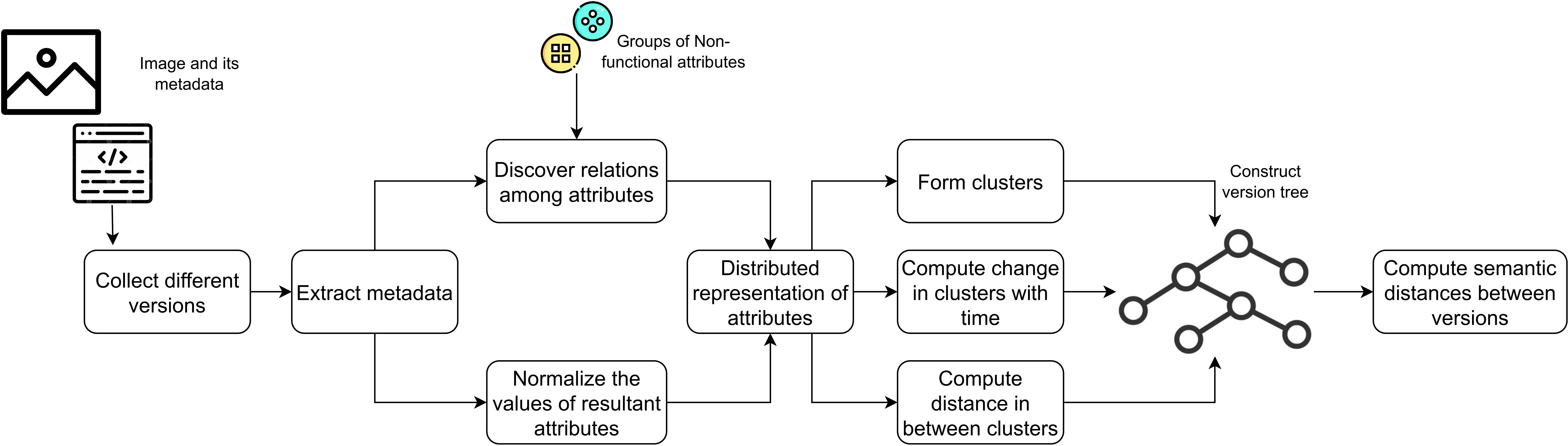}}\caption{The Proposed Framework}
\label{fig:framework}
\vspace{-2em}
\end{figure}

This Section provides details of our novel framework (shown in Figure~\ref{fig:framework}).

\subsection{Feature Extraction}
The proposed framework takes non-functional attributes of an image service as an input and determines whether the image is modified. 
The first step is to extract the meta-information available with the image which resides in the metadata of the uploaded image and the information posted with the image. We propose to perform image provenance analysis using the metadata of different versions. In this regard, we assume that the provided image is not the only version available on social media. Therefore, we utilize \textit{Reverse Image Search} to collect all versions of the provided image along with their metadata. 

\subsection{Grouping the Non-functional Attributes}
 
We propose to determine changes in an image service using only the non-functional attributes, i.e., without using the image itself. The non-functional attributes should ideally be completely reflective of the content in the picture to correctly reflect on the changes in an image. However, state-of-the-art representation of non-functional attributes is not well-reflective of the semantics of an image. 
Therefore, we propose a novel representation of these non-functional attributes in form of different groups in Table~\ref{table:groups_Nonfunctional_Attributes}. 
The attributes are placed in a same group if observing them collectively can provide some useful insights about the changes.
Analyzing each group reveals useful insights about the picture. These insights are mostly related to spatio-temporal and contextual parameters of an image service.
For instance, \textit{shutter speed} and \textit{exposure time} are placed in a same group because analyzing these attributes collectively indicates the intake of light while capturing an image. It is indirectly reflective of the \textit{time of day} when the picture was captured.

\begin{table}
\vspace{-2em}
\caption{Grouping of Non-functional Attributes}\label{table:groups_Nonfunctional_Attributes}
\begin{tabular}{|p{4cm}|p{8cm}|}
\hline
\textbf{Attributes} & \textbf{Description}\\
\hline
\multirow{2}{4cm}{DateTimeOriginal, DateTimeDigitized, DateTime}  
& \multirow{3}{8cm}{An image metadata contains various timestamps. By comparing these timestamps, we can see whether the image has been modified.} \\ &  \\&  \\ &\\
  
\multirow{2}{4cm}{Make, Model, Other attributes (focal length, aperture, etc.)}  
& \multirow{3}{8cm}{A camera's make and model can be compared with different attributes i.e., focal length, aperture size, and resolution etc. to check whether the camera supports these attribute values.} \\ &  \\  &  \\&  \\

\multirow{3}{4cm}{Aperture, Shutter speed, Exposure value} 
& \multirow{4}{8cm}{Exposure value is calculated from aperture and shutter speed. We can re-compute exposure value and compare it with the attributes.} \\ & \\ & \\ & \\

\multirow{4}{4cm}{Shutter speed, Aperture, ISO} 
& \multirow{5}{8cm}{The three pillars of exposure are shutter speed, aperture, and ISO. These three attributes are interconnected and changing one affects the others. 
For example, increasing the shutter speed may require a wider aperture or higher ISO to compensate for the reduced amount of light reaching the sensor. Similarly, using a narrow aperture for a deeper depth of field may require a slower shutter speed or higher ISO to compensate for the reduced amount of light.} \\ & \\ & \\ & \\ & \\ & \\ & \\ & \\ &\\ &\\

\multirow{2}{4cm}{Aperture, Exposure Time, Shutter speed, Datetime} 
& \multirow{3}{8cm}{Exposure time and shutter speed reflects the light intake in a picture. It can be transformed to the time of day based on the following: Noon: high f/stop, fast shutter speed; Night: low f/stop, slow shutter speed to let more light in. However, it depends on the environment (indoor/outdoor) and purpose of shooting.
} \\ & \\ & \\ & \\ & \\ &\\ &\\

\multirow{3}{4cm}{GPS info, TimeZoneOffset, DateTime, Location} 
& \multirow{3}{8cm}{We analyze these attributes to see whether the GPS is consistent with the timezone. Moreover, we check whether the datetime is consistent with the timezone.} \\ & \\ & \\ &\\

\multirow{2}{4cm}{Datetime, GPS info, ISO, white balance and other camera settings} 
& \multirow{3}{8cm}{By using the datetime and GPS info, we can find weather conditions of the shooting location. The white balance and ISO depends on the weather. 
} \\ & \\ & \\ & \\

\multirow{2}{4cm}{Temperature, Humidity, Pressure, Weather} 
& \multirow{2}{8cm}{Using GPS and timestamp, we can retrieve weather information of location of shooting using online APIs. 
} \\ & \\ & \\

\multirow{3}{4cm}{WaterDepth, GPS info} 
& \multirow{3}{8cm}{Using GPS info, we can get the water depth information of the shooting location that can be compared with the given WaterDepth to see if it is consistent.} \\ & \\ & \\ & \\

\hline
\end{tabular}
\vspace{-2em}
\end{table}

\subsection{Normalizing the Non-functional Attributes}
Attributes in a group may have different units and scales.
The attributes in each group are first normalized on a common scale. In this regard, we transform information provided by each group shown in Table~\ref{table:groups_Nonfunctional_Attributes} in to their normalized values. For instance, if we have three attributes i.e., $\zeta$, $\tau$ and $c$, then these attributes can be normalized using the following equation:

\begin{equation}
\hat{\zeta} = \frac{\zeta _i - min(\zeta, \tau, c)}{max(\zeta, \tau, c) - min(\zeta, \tau, c)}
\end{equation}
where $\hat{\zeta}$ represents the normalized value of $\zeta$.
$\tau$ and $c$ can be normalized similarly.

\subsection{Creating a Distributed Representation of Attributes}
\begin{figure}
\vspace{-1.5em}
\centering
\begin{minipage}{.4\textwidth}
  \centering
  \includegraphics[width=\linewidth]{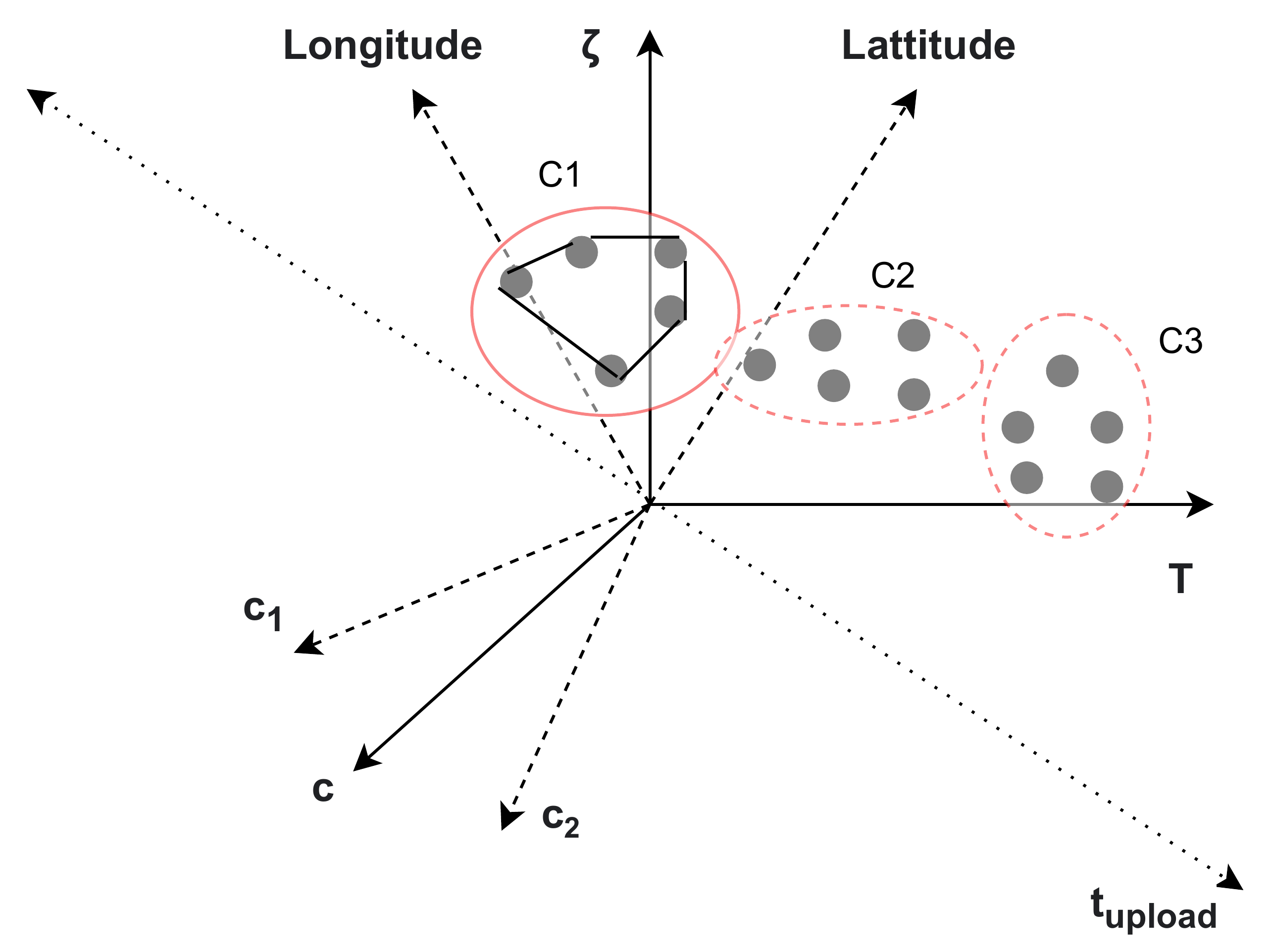}
  \caption{Distributed Representation of Attributes}
  \label{fig:distributed_space}
\end{minipage}%
\begin{minipage}{.4\textwidth}
  \centering
  \includegraphics[width=\linewidth]{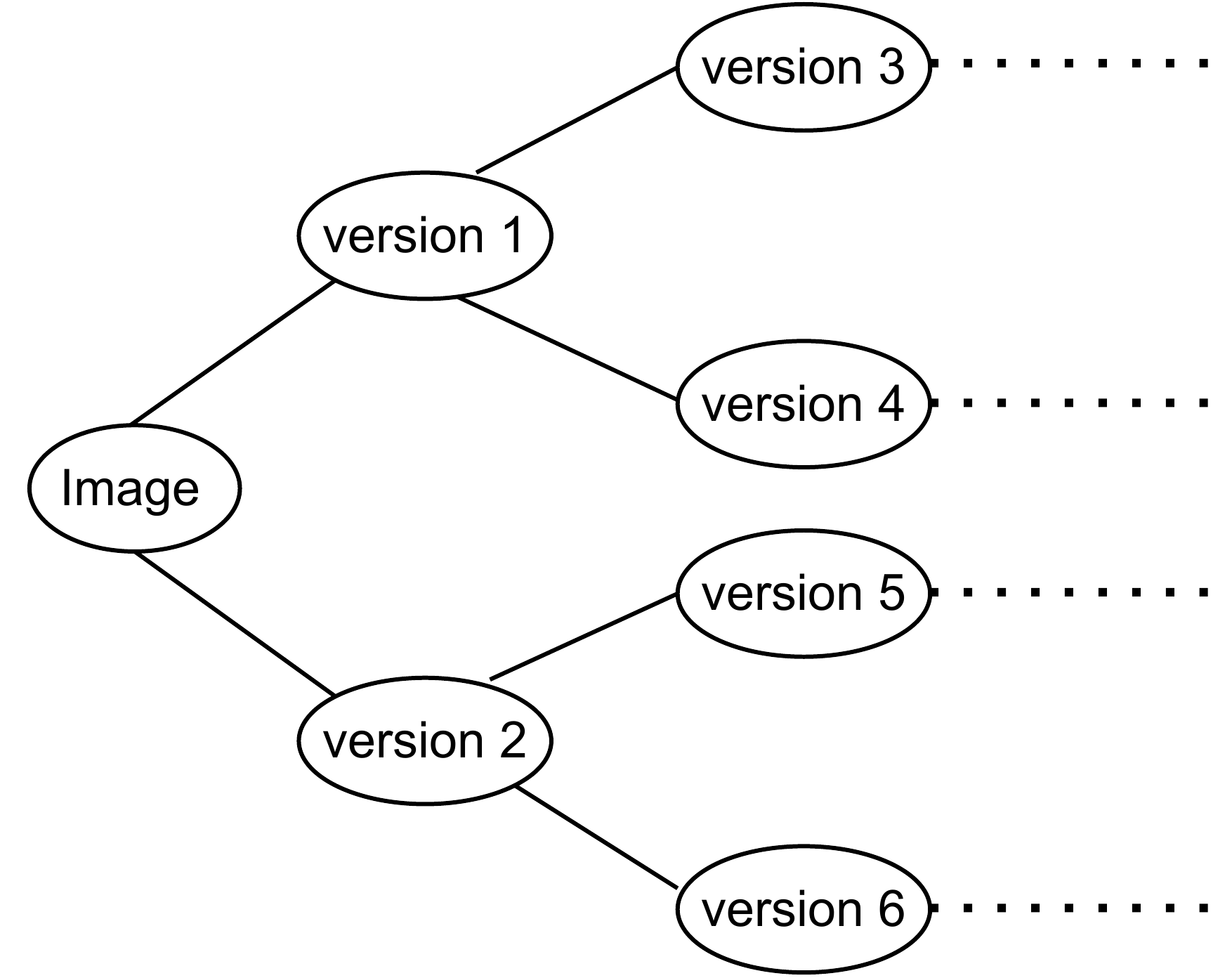}
  \caption{Version Tree}
  \label{fig:versionTree}
\end{minipage}
\vspace{-2em}
\end{figure}

We create a distributed representation of the normalized non-functional attributes. Each attribute is represented on a high dimensional space. The number of dimensions depends on the types of attributes in a group. In most cases, spatio-temporal and contextual attributes constitute the Cartesian space. These dimensions are further composed of multiple sub-dimensions. For instance, the spatial axis is represented by two axes: longitude and latitude. Similarly, a contextual attribute can be represented as a multi-dimensional vector using either Latent Semantic Analysis or any Word-Embeddings-based approach. Therefore, the contextual axis is further divided in to multiple sub-axes as shown in Figure~\ref{fig:distributed_space}. Afterwards, we plot the values of each group in this high dimensional space. 
We then cluster the attribute values of each group for a specific version.
The same process is repeated for each group. As a result, we get different clusters plotted in this distributed space as shown in Figure~\ref{fig:distributed_space}. Each cluster corresponds to a single version of an image.
It is worth mentioning that there is an additional time axis ($t_{upload}$) in the distributed space that informs the time of upload of a specific version. Due to this axis, clusters cannot be overlapped completely. For two clusters to overlap, their versions should be uploaded at exactly the same time. The clusters projected in this space provides a holistic view of non-functional attributes of all versions. The differences in placements and shapes of clusters reflect discrepancies among the versions. Moreover, on time axis, it also reflects the evolution of different versions from an image. However, this sequence of uploads may not accurately inform about the sequence of changes introduced in an image. For instance, version corresponding to cluster C3 in Figure~\ref{fig:distributed_space} is uploaded after C2, however, in reality, C3 may be the predecessor of C2. To address it, we transform the sequence of uploads to the sequence of changes. 

\subsection{Version Tree}
Most viral fake social media images are usually related to a specific incident i.e., a road accident, a natural disaster, a man-made disaster etc. Manipulating facts about these incidents may involve sharing relevant fake posts. We consider \textit{relevant images} and \textit{versions of an image} to be two different entities. Relevant images contain relevant stuff but they are not originated from a same image. Whereas, an image $b$ is a version of another image $a$ if $b$ is originated from $a$. It is crucial to analyze different versions of an image to detect changes in an image service because it provides a holistic view of manipulations about a specific incident. We propose an image provenance analysis based only on the image metadata. The provenance analysis results in a tree because of the fact that an image version can only be originated from one single image. Therefore, we propose to construct a version tree which is reflective of the modifications introduced in each version of an image. It also informs about the transformation of different attributes in that image. Figure~\ref{fig:versionTree} shows a generic version tree. 

\noindent\textit{\textbf{Definition 1:}}
An \textit{image b} is assumed to be version of another \textit{image a} if 
\begin{itemize}
    \item \textit{Image b} originated from \textit{image a}.
    
    \item Attributes in \textit{image a} can be transformed to attributes in \textit{image b}. We leverage the concept of transformation matrix to analyze the nature of transform. 
    Equation~\ref{eq: Matrix} shows the linear transformations of one cluster to another.
    
    \begin{equation}
    \Delta_t \times I_a = I_b 
\: where \:
    \Delta_t =
    \begin{array}{cc}
    \begin{bmatrix}
                     w&x&.\\
                     y&z&.\\
                     .&.&.
    \end{bmatrix}
    \end{array}
\: and \:
    I_a =
    \begin{array}{c}
    \begin{bmatrix}
                     attr_1\\
                     attr_2\\
                     .
    \end{bmatrix}
    \end{array}
    \label{eq: Matrix}
    \end{equation}
 
    where $attr$ stands for attribute, $I_a$ is an attribute matrix for image a, $I_b$ is an attribute matrix for image b, and $\Delta_t$ is a linear transformation matrix. According to this statement, even if two images $image_b$ and $image_c$ are originated from $image_a$, if $\Delta_t \times I_b = I_c$ is satisfied, then $image_c$ will be considered as a version of $image_b$.
    
    \item Linear transformation can cover the following transformations: translation, scaling, rotation, reflection, shearing, projection, orthogonal projection and affine transformation. If the relationship between the two clusters is inherently nonlinear, a quadratic transformation (shown in equation~\ref{eq: version}) may better capture the underlying transformation. Therefore, we consider the matrix form of quadratic transformations if the linear transformation of an attribute matrix can not be found or is too complex. The criteria to select linear or quadratic transformation depends on the complexity of the transformation. We choose the transform with a simpler transformation matrix.
    
    \begin{equation}
    I_a^t \times \Delta_t \times I_a = Quadratic \: form \: of \: I_b 
    \label{eq: version}
    \end{equation}
    
\end{itemize}

    We consider the complexity of $\Delta_t$ to check if $\Delta_t \times I_a = I_b$ is satisfied. The lower the complexity, higher will be the likelihood for $image_b$ to be version of $image_a$. The complexity is determined by the similarity of the transformation matrix with the identity matrix (the matrix that, if used as a transformation matrix, returns the same input matrix).
    However, to determine the complexity of quadratic transformation, we compute the difference of the transformation matrix with the matrix that, if used as a transformation matrix, returns quadratic form of the input matrix. It is worth mentioning that the tree proposed in Figure~\ref{fig:versionTree} is implicitly directional. In the process of constructing the version tree, it automatically determines the origin of the image. However, in case of a linear tree i.e., if there is a single path from root to a leaf node, then the direction of the tree can be deduced from the order of uploads of images.

The version tree provides a holistic view of the changes introduced in different versions of an image. More importantly, it informs about the transformation of an image in to different versions that serves as a tool for image provenance analysis. The depth of each node in the tree is reflective of the extent of changes in the corresponding version. These insights can not be reflected by reporting changes in each version individually.

\subsection{Heuristics to Construct a Version Tree}
We propose novel heuristics to construct version tree for an image. 
We utilize the clustering and the transformation matrix proposed in the previous Sections to construct the version tree. 
The clusters projected on the Cartesian space are the groups of attributes of different versions. Each cluster corresponds to a specific group of a specific version.
The distance between these clusters reflect the changes among versions. The variations in the clusters actually represent the transformation of a version in to another. As stated earlier, we consider the sequence of uploads of image versions as a baseline solution. Therefore, we propose reordering of the clusters on the time axes to make it accurate. It will result in a sequence of edits/updates in an image. 
The criteria to start the swapping is to check whether attributes in a version are being transformed from another version. To ensure it, we rely on the transformation matrix proposed in equation~\ref{eq: version}. According to equation~\ref{eq: version}, there will be multiple transformation matrices for each version because we have a transformation matrix for each group of non-functional attributes. Therefore, for each version, we have $\Delta_1$ to $\Delta_n$ matrices where n represents the total number of groups. In the next step, we find the nearest neighbors for each $\Delta$ using the Frobenius Distance:
    \begin{equation}
     F_{a,b} = \sqrt{trace((a-b) \times (a-b)')} 
     \label{eq:first_criteria}
    \end{equation}
Next, we need to decide the sequence of versions, $V_1$ is placed before $V_2$ if $\Delta_1$ to $\Delta_n$ of $V_2$ can be derived using either linear or quadratic transformation.
The reordering is only possible if equation~\ref{eq: Matrix} or \ref{eq: version} is satisfied for two consecutive versions on a time axis. Another factor, on which we rely to decide about swapping is the similarity between two versions. If the clusters shapes are similar for two distant versions, we do reordering to make them closer. The following equation is used to check this criterion:
    \begin{equation}
     if \quad C_{n+1} \cap C_n \geq C_n \cap C_{n-1}
     \label{eq:similarity_criteria}
    \end{equation}

We need a control statement to decide how many iterations are required to reach an optimal sequence of versions. 
In this regard, we introduce a threshold in equation~\ref{eq:similarity_criteria} to control the reordering:

    \begin{equation}
     if \quad (C_{n+1} \cap C_n) - (C_n \cap C_{n-1}) \geq Threshold
     \label{eq:similarity_criteria}
    \end{equation}

The value of threshold is assigned in a way that the criterion for reordering becomes relatively strict after every swap.  
\begin{figure}
\centering
\begin{minipage}{.35\textwidth}
  \centering
  \includegraphics[width=\linewidth]{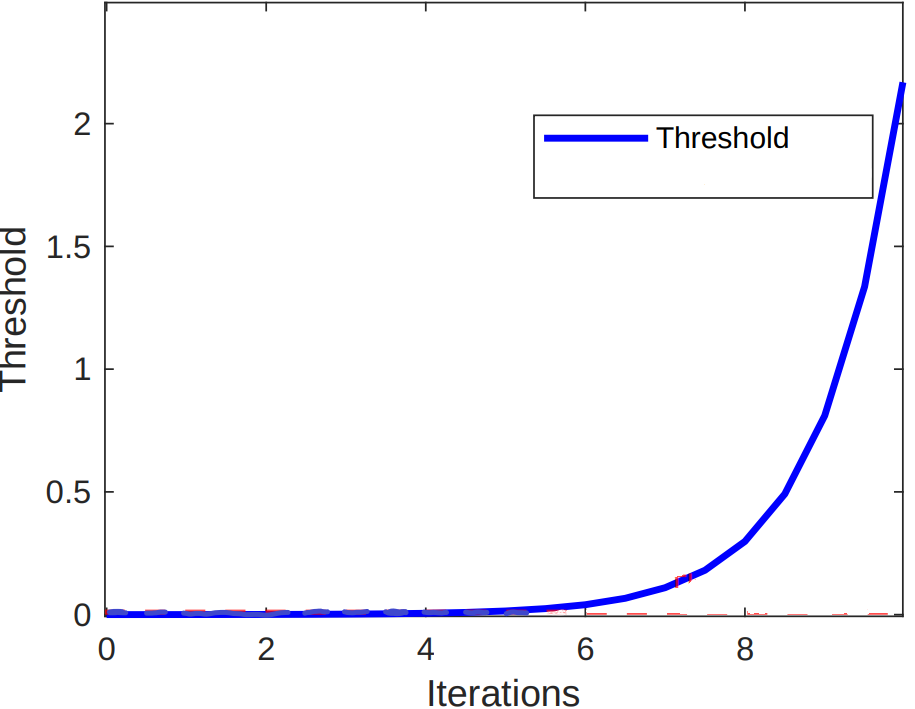}
  \caption{Trend of Threshold}
  \vspace{-0.5em}
  \label{fig:threshold}
\end{minipage}%
\begin{minipage}{.35\textwidth}
  \centering
  \includegraphics[width=\linewidth]{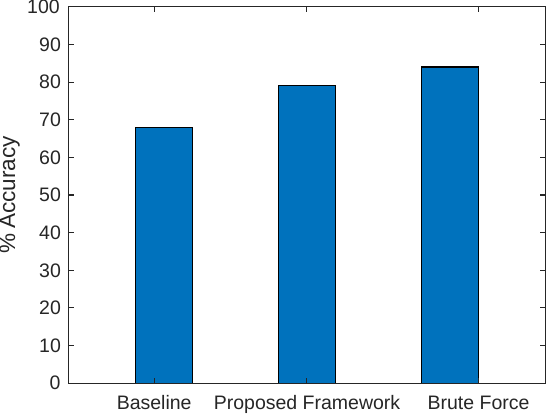}
  \caption{Accuracy}
  \vspace{-0.5em}
  \label{fig:Accuracy}
\end{minipage}
\end{figure}
As evident from Figure~\ref{fig:threshold}, the threshold follows an exponential trend. The exponential trend is reflective of the fact that the reordering becomes relatively less likely after each iteration.
However, the difference/increase in threshold is not linear. Therefore, the change in threshold must be computed after each interval. We calculate this gap based on the similarity of changes between two consecutive clusters/versions on the time axis. The less similar the changes are in two versions, the more likely should be the swap, so we can assign a higher increase in threshold. The idea about the design of the threshold is presented in~\cite{umair2020energy,umair2023energy}.

\begin{equation}
     \Delta Threshold \propto (\Delta C_n - \Delta C_{n-1})
     \label{eq:increaseinThreshold}
\end{equation}

This procedure constructs one branch of the version tree. To construct other branches, the same procedure is repeated for the remaining nodes/versions. Once a version tree is constructed, the consecutive version nodes can be compared using any semantic similarity measure to find its semantic difference with the previous version. 
Comparing different nodes in the version tree results in $\delta \zeta$, $\delta \tau$ and $\delta c$, where $\delta$ represents a change. The proposed framework is self-configurable. Whenever, it encounters a distinct type of version, it becomes part of its learning. The concept of self-configurable algorithms is presented in~\cite{umair2018self}.


\section{Experimentation and Results}\label{sec:experiments}
\begin{figure}
\vspace{-1.5em}
\centerline{\includegraphics[width = 1\columnwidth]{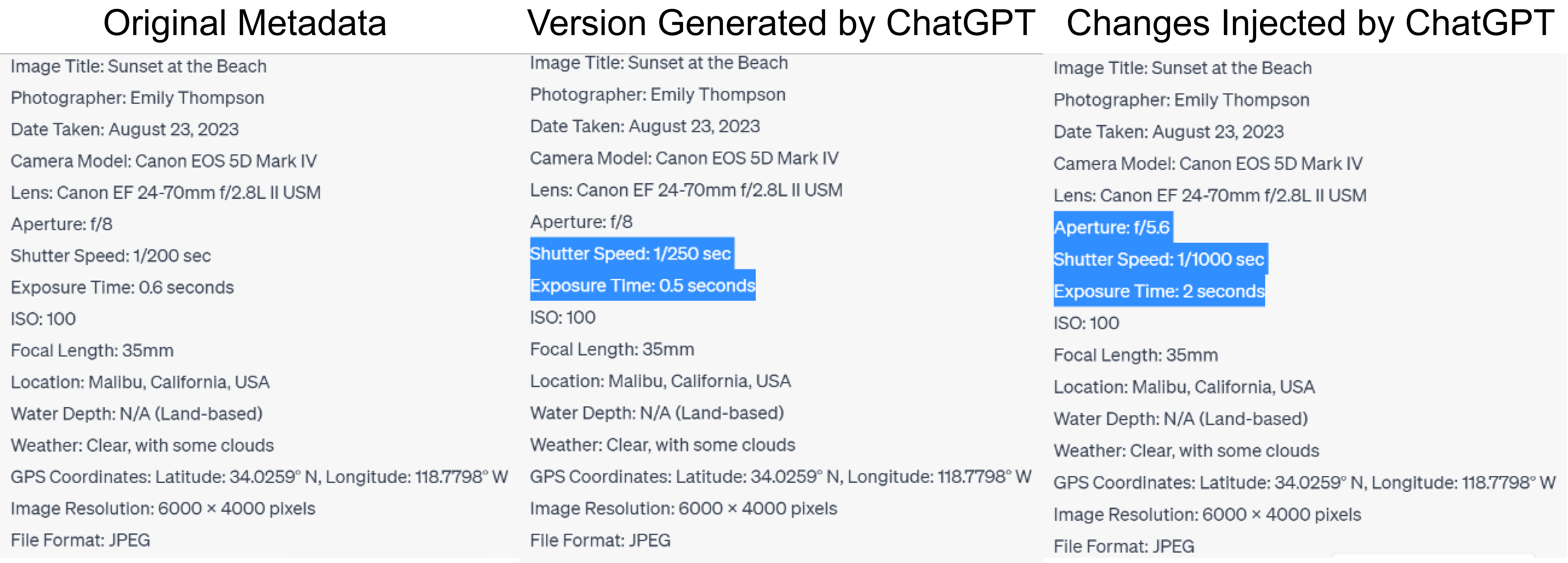}}\caption{A Sample Metadata and Changes Introduced by ChatGPT}
\vspace{-2.95em}
\label{fig:MetadataChatGPT}
\end{figure}

We use a real image metadata dataset to conduct experiments~\cite{drewnoakes2020}. The dataset contains images and videos along with their metadata. The dataset includes a wide range of images covering various subjects, scenes, and visual characteristics. We use the metadata-extractor library, which is a Java-based library for reading metadata from image files. The extracted metadata encompasses information such as camera make and model, image dimensions, capture date and time, GPS coordinates, and other technical details. Although, the dataset provides images, however, we only exploit their metadata. We use ChatGPT to generate different versions of images contained in this metadata dataset. ChatGPT can be effectively utilized to generate versions of image metadata due to its language generation capabilities and understanding of contextual information. We also leverage ChatGPT to introduce systematic changes in the sample metadata of different image versions. Instructions are provided to the ChatGPT to create different types of variations in between metadata of different versions of an image. For instance, in some images, shutter speed and exposure time have been made inconsistent. By leveraging its language generation capabilities, ChatGPT can produce altered metadata such as updated timestamps, modified camera settings, or edited descriptions, providing an indication that the image has undergone changes. Figure~\ref{fig:MetadataChatGPT} shows a sample metadata of two versions generated by ChatGPT. It also shows inconsistencies introduced by ChatGPT. The inconsistencies are shown in highlighted text in Figure~\ref{fig:MetadataChatGPT}.

The experiments are completely scalable as we use an API provided by OpenAI to execute text commands on ChatGPT. 
The API allows us to make requests to the ChatGPT model hosted on OpenAI's servers and receive responses in real-time, enabling interactive and dynamic conversations with the language model. By utilizing the API, we leverage OpenAI's infrastructure to handle the computational resources required for generating versions of an image metadata, ensuring scalability and availability. Moreover, the API provides quick responses, allowing for real-time interactions and dynamic conversations with ChatGPT. We use the HTTP POST method to send a request to the API endpoint. We structure the request payload in JSON format. The payload may contain a list of message objects with a role (either "system", "user", or "assistant") and content (the text of the message).
We cluster the attributes of each version for each group. We use the clustering approach proposed in~\cite{rizvi2021clustering}.
The distance between the clusters and the dissimilarities among them are leveraged to determine the transformation matrix to build the version tree.

\subsubsection{Effectiveness}
We report the performance of the proposed approach in terms of accuracy (in Figure~\ref{fig:Accuracy}) and run-time. The ground truth of metadata is known from the metadata dataset.
Accuracy is calculated as the percentage of correctly classified image versions. We compare our approach with a baseline and a brute force approach. Sequence of uploads of an image (the order in which the image versions were originally uploaded) on social media is regarded as the baseline solution, whereas, the brute force approach doesn't consider the criteria defined in equation~~\ref{eq:similarity_criteria} and \ref{eq:increaseinThreshold}. The proposed framework achieves an accuracy of 76\%. Whereas, the baseline approach is correct 64\% of the times. Brute force performs better than the proposed framework but at the cost of additional computations. We also report the run-time complexity and the time consumed (in nano seconds) for brute force and our proposed approach as shown in Table~\ref{tab:timeEff}. Time complexity is computed by counting units of time. Moreover, the increase in run-time with the increase in the number of inconsistencies is reported in Figure~\ref{fig:timeEff}.

\begin{table}[htbp]
\vspace{-3em}
\caption{Run Time Efficiency}
\vspace{-1em}
\begin{center}
\begin{tabular}{|c|c|c|c|}
\hline
&\textbf{Baseline} & \textbf{Brute-force} & \textbf{Heuristics} \\ \hline
\textbf{Run-time Complexity} & 1 & O(N$^2$) & O(N$^{1/2}$) \\
\textbf{Time Consumed (ns)} & 309 & 26500 & 15300 \\
\hline
 
\end{tabular}
\vspace{-3.5em}
\label{tab:timeEff}
\end{center}
\end{table}

\subsubsection{Comparison}
\begin{figure}
\vspace{-2em}
\centering
\begin{minipage}{.4\textwidth}
  \centering
  \includegraphics[width=\linewidth]{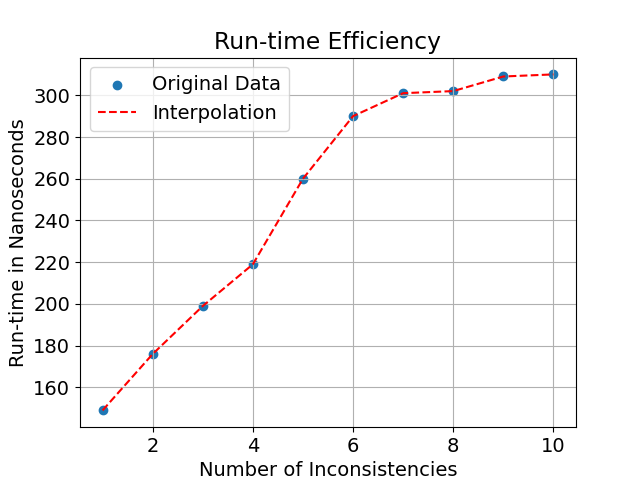}
  \caption{Run-time Efficiency}
  \vspace{-3em}
  \label{fig:timeEff}
\end{minipage}%
\hspace{1cm}
\begin{minipage}{.35\textwidth}
  \vspace{0.3em}
  \centering
  \includegraphics[width=\linewidth]{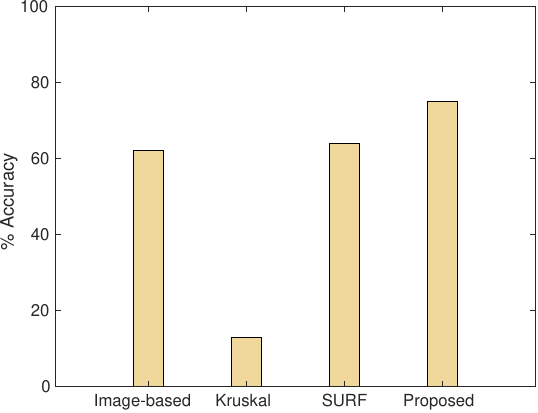}
  \caption{Comparison with State-of-the-art}
  \vspace{-3em}
  \label{fig:SOTA}
\end{minipage}
\end{figure}

We compare the accuracy of our proposed framework with a state-of-the-art that uses image metadata along with image content to perform image provenance analysis~\cite{bharati2019beyond}. The framework proposed in~\cite{bharati2019beyond} has three variants: a complete image-based solution; Kruskal's maximum spanning tree algorithm based only on image metadata; and Cluster-SURF which utilizes both images and their metadata. The accuracy of the approach is computed in terms of the overlap between the original version tree and the constructed version tree. Our proposed approach outperforms all variants as reflected in Figure~\ref{fig:SOTA}. 

\section{Conclusion}\label{sec:conclusion}
We propose a novel framework to detect changes in an image service using only the non-functional attributes. 
The proposed model returns a version tree for an image service. 
Theory of matrix transformation is leveraged in this paper to model the transformation of one image version to another. The results validate the proposed approach. This work can be further extended to investigate the detected changes to check whether the changes are constituting a fake. One aspect to consider is that run-time may increase with a larger scale of images.


%
%

%
%
%
 \bibliographystyle{splncs04}
 \bibliography{ref}

\end{document}